\documentclass[12pt]{article}

\newcommand{\be}{\begin{equation}}
\newcommand{\ee}{\end{equation}}
\newcommand{\bi}[1]{\vspace{-3mm} \bibitem{#1}}
\usepackage{epsfig,amsmath,amssymb,graphics,graphicx} 

\usepackage{amscd}
\usepackage{pb-diagram}
\usepackage{hyperref}
\usepackage{amsfonts}

\begin{document}

\begin{center}

{\bf \large Fractional Derivative Regularization in QFT} 

\vskip 7mm
{\bf \large Vasily E. Tarasov} \\
\vskip 3mm

{\it Skobeltsyn Institute of Nuclear Physics,\\ 
Lomonosov Moscow State University, \\
Moscow 119991, Russia} \\
{E-mail: tarasov@theory.sinp.msu.ru} \\

\begin{abstract}
In this paper, we propose new regularization, 
where integer-order differential operators are 
replaced by fractional-order operators.
Regularization for quantum field theories
based on application of 
the Riesz fractional derivatives of 
non-integer orders is suggested.
The regularized loop integrals depend on parameter
that is the order $\alpha>0$ of the fractional derivative.
The regularization procedure is demonstrated
for scalar massless fields in $\varphi^4$-theory 
on $n$-dimensional pseudo-Euclidean space-time.
\end{abstract}

\end{center}


\section{Introduction}

A characteristic feature of quantum field theories, 
which are used in high-energy physics, 
is ultraviolet divergence \cite{ Bogol-Shirk,Ryder}.
These divergences arise in momentum space 
from modes of very high moment, i.e. 
the structure of the field theories at very short distances.
In the narrow class of quantum theories,
which are called "renormalizable", 
the divergences can be removed 
by a singular redefinition of the parameters of the theory. 
This process is called the renormalization \cite{Collins}, 
and it defines a quantum field theory 
as a non-trivial limit of theory with a ultraviolet cut-off.

The renormalization requires the regularization of 
the loop integrals in momentum space.
These regularized integrals depend on parameters
such as the momentum cut-off, the Pauli-Villars masses, 
the dimensional regularization parameter, 
which are used in the corresponding regularization procedure. 
This regularized integration is ultraviolet finite. 

We suggest a regularization procedure 
based on fractional-order derivatives of the Riesz type.
In momentum space these derivatives are represented
by power functions of momentum.



Fractional calculus and fractional differential equations
\cite{SKM,KST} 
have a wide application in mechanics and physics.
The theory of integro-differential equations of 
non-integer orders is a powerful tool to describe 
the dynamics of systems and processes with power-law 
non-locality, long-range memory and/or fractal properties.

Recently the spatial fractional-order derivatives 
have been actively used in 
the space-fractional quantum mechanics \cite{Tar-1,Tar-5},
the quantum field theory and gravity 
for fractional space-time \cite{Calcagni1},
the fractional quantum field theory 
at positive temperature \cite{Lim-1,Lim-3}.

Fractional calculus allows us to take into account
fractional power-law nonlocality of 
continuously distributed systems and classical fields.
Using the fractional calculus, we can consider
space-time fractional differential equations 
in the quantum field theory. 
For the first time, the fractional-order Laplace and d'Alembert 
operators have been suggested by Riesz in \cite{Riesz}.
Then non-integer powers of d'Alembertian 
are considered in different works
(for example, see Section 28 in \cite{SKM} and 
\cite{Dal-1,Dal-2,Dal-3}).
The fractional Laplace and d'Alembert operators
of the Riesz type are the basis for
the fractional field theory in multidimensional spaces.

As it was shown in \cite{JPA2006,JMP2006,JPA2014,AMC2015,CMA2017}, 
the continuum equations with fractional derivatives 
of the Riesz type can be directly connected to
lattice models with long-range properties. 
A connection between the dynamics of lattice system 
with long-range properties 
and the fractional continuum equations are proved 
by using the transform operation 
\cite{JPA2006,JMP2006,JPA2014,AMC2015}.

There are different definitions of fractional derivatives
such as Riemann-Liouville, Caputo, Gr\"unwald-Letnikov, 
Marchaud, Weyl, Sonin-Letnikov, Riesz \cite{SKM,KST}.
Unfortunately all these fractional derivatives 
have a lot of unusual properties.
For example, the well-known Leibniz rule 
does not hold for differentiation of non-integer orders 
\cite{CNSNS2013}.

It should be noted that the use of the fractional derivative of 
non-integer order is actually equivalent to using 
an infinite number of derivatives of all integer orders. 
For example, the Riemann-Liouville derivative 
$\, ^{RL}{\cal D}^{\alpha}_{a+}$
can be represented in the form of the infinite series 
\be \label{UUP-4}
(\, ^{RL}{\cal D}^{\alpha}_{a+} f)(x) = \sum^{\infty}_{k=0} 
\frac{\Gamma(\alpha+1)}{\Gamma(k+1) \Gamma(\alpha-k+1)} 
\frac{(x-a)^{k-\alpha}}{\Gamma(k-\alpha+1)} 
\frac{d^k f(x)}{dx^k} \ee
for analytic (expandable in a power series on the interval) functions on $(a,b)$,
(see Lemma 15.3 of \cite{SKM}).
Therefore, an application of fractional derivatives means 
that we take into account a contribution of derivatives 
of all integer orders with the power-law weight.

In the suggested FD-regularization we use fractional derivatives 
of non-integer orders $\alpha$ close to integer values.
The use of these fractional derivatives 
of non-integer orders $\alpha$ close to integer values
in the regularization procedure means that we
consider a weak nonlocality,
i.e. a slight deviation from the locality. 


\section{Fractional Laplacian 
for Euclidean space $\mathbb{R}^n$}


\subsection{Fractional integration in the Riesz form}

The Riesz fractional integral of order $\alpha$ 
for $\mathbb{R}^n$ is defined \cite{Riesz} (see also 
Section 25 of \cite{SKM}) by equation
\be
^RI^{\alpha} f(P) =
\frac{1}{H_n (\alpha)} \int_{\mathbb{R}^n}
f(Q) \,r^{\alpha-n}_{PQ} dQ ,
\ee
where $P$ and $Q$ are points of the space $\mathbb{R}^n$, and
\be \label{HnE}
H_n(\alpha) = 
\frac{\pi^{n/2} 2^{\alpha} \Gamma (\alpha/2)}{
\Gamma ((n-\alpha)/2)} .
\ee

Let us give some properties of the Riesz
fractional integral that are proved in \cite{Riesz}:

1. The semi-group property of the Riesz fractional integration
\be \label{SemiGr}
^RI^{\alpha_1} \, ( \,^RI^{\alpha_2} f(P) ) =
\, ^RI^{\alpha_1+\alpha_2} f(P) ,
\ee
where $\alpha_1 >0$, $\alpha_2 >0$, $\alpha_1 +\alpha_2 <n$.

2. The action of the Laplace operator on 
the Riesz fractional integral 
\be
\Delta \, ^RI^{\alpha+2} f(P) = - \, ^RI^{\alpha} f(P) ,
\ee
and we also have 
\be \label{DI=I-1}
\Delta^k \, ^RI^{\alpha+2k} f(P) = (-1)^k 
\, ^{R}I^{\alpha} f(P) , \quad (k \in \mathbb{N}).
\ee

3. The Riesz fractional integration of exponential function
does not change this function
\be
^RI^{\alpha} \, e^{i x_j} = e^{i x_j} ,
\ee
\be
^RI^{\alpha} \, e^{i \sum^n_{j=1} a_j x_j} =
\left(\sum^n_{j=1} a_j x_j \right)^{- \alpha/2} 
\, e^{i \sum^n_{j=1} a_j x_j} .
\ee

There is the following important property
\be \label{Tru}
^RI^{\alpha} f(P) = (-1)^m \,
^RI^{\alpha+2m} \, \Delta^m f(P) .
\ee
This property allows us to use an analytic continuation of 
$^RI^{\alpha} f(P)$ for negative values of 
$\alpha > - 2m$, where $m \in \mathbb{R}$ (see \cite{Riesz}).
In this case, the semigroup property (\ref{SemiGr}) 
can be used for $\alpha_1 > - 2m$, $\alpha_2 > - 2m$, 
$\alpha_1+ \alpha_2 > - 2m$, and $m \in \mathbb{R}$.
Analogously, we can consider the property 
(\ref{DI=I-1}) in the form
\be \label{DI=I-2}
\Delta^k \, ^RI^{\alpha} f(P) = (-1)^k 
\, ^{R}I^{\alpha-2k} f(P) , \quad (k \in \mathbb{N}) .
\ee


\subsection{Fractional Laplacian in the Riesz form}

For the first time  fractional Laplace operators 
have been suggested by Riesz in \cite{Riesz}
(see also Section 25 of \cite{SKM}).
The fractional Laplacian $(- \Delta)^{\alpha/2}$ 
in the Riesz form for $n$-dimensional Euclidean 
space $\mathbb{R}^n$ can be considered \cite{SKM}
as an inverse Fourier's integral transform 
${\cal F}^{-1}$ of $|{\bf k}|^{\alpha}$ by 
\be \label{RFD-1}
( (- \Delta)^{\alpha/2} \varphi)(x)= 
{\cal F}^{-1} \Bigl( |{\bf k}|^{\alpha} ({\cal F} \varphi)
({\bf k}) \Bigr) ,
\ee
where $\alpha > 0$ and $x \in \mathbb{R}^n$. 

For $\alpha >0$, the fractional Laplacian of 
the Riesz form can be defined \cite{SKM}
as the hypersingular integral
\be \label{RFD-2}
\left( (- \Delta)^{\alpha/2} \varphi \right)(x) =
\frac{1}{d_n(m,\alpha)} \int_{\mathbb{R}^n} 
\frac{1}{|z|^{\alpha+n}} 
(\Delta^m_z \varphi) (z) \, d^4 z ,
\ee
where $m> \alpha$, and $(\Delta^m_z \varphi)(z)$ 
is a finite difference of order $m$ of a field 
$\varphi (x)$ with
a vector step $z \in \mathbb{R}^n$
and centered at the point $x \in \mathbb{R}^n$:
\[ 
(\Delta^m_z \varphi)(z)
= \sum^m_{j=0} (-1)^j \frac{m!}{j!(m-j)!} \, 
\varphi (x-jz) . 
\]
The constant $d_n(m,\alpha)$ is defined by
\[ 
d_n(m,\alpha)=\frac{\pi^{1+n/2} A_m(\alpha)}{2^{\alpha} 
\Gamma(1+\alpha/2) \Gamma((n+\alpha)/2) \sin (\pi \alpha/2)} , 
\]
where
\[ 
A_m(\alpha)=
\sum^m_{j=0} (-1)^{j-1} \frac{m!}{j!(m-j)!} \, j^{\alpha} . 
\]

Note that the hypersingular integral (\ref{RFD-2})
does not depend on the choice of $m>\alpha$.
The Fourier transform ${\cal F}$ of 
the fractional Laplacian is given by 
${\cal F} \{(- \Delta)^{\alpha/2} \varphi \}({\bf k}) = 
|{\bf k}|^{\alpha} ({\cal F} \varphi)({\bf k})$. 
This equation is valid for the Lizorkin space \cite{SKM}
and the space $C^{\infty}(\mathbb{R}^n)$ 
of infinitely differentiable 
functions on $\mathbb{R}^n$ with compact support.

\subsection{Fractional Laplacian in the Riesz-Trujillo form}

Using  the property (\ref{Tru}), 
we can define the fractional Laplace operator by the equation
\be
^{RT}\Delta^{\alpha/2} f(P) = (-1)^m \,
^RI^{2m-\alpha} \, \Delta^m f(P) ,
\ee
where $m > \alpha/2$ and $m \in \mathbb{R}$.
This form of definition of the fractional Laplacian
has been suggested by J.J. Trujillo in 2012 
(see also \cite{PRTV}).

It is important that the semi-group property holds
for these operators
\be \label{SemiGr-LP}
^{RT}\Delta^{\alpha_1/2} \,
^{RT}\Delta^{\alpha_2/2} f(P) = \,
^{RT}\Delta^{(\alpha_1+\alpha_2)/2} f(P) ,
\ee
where $0<\alpha_1 <2m$, $0<\alpha_2 <2m$, 
$\alpha_1 +\alpha_2 < 2m$.

We can see that
\be
^{RT}\Delta^{k} f(P) = (-1)^k \, \Delta^k f(P) ,
\ee
where $k \in \mathbb{N}$.
In the Riesz's notation \cite{Riesz},
this equation means that
\be
^{R}I^{-2k} f(P) = (-1)^k \, \Delta^k f(P) .
\ee
We also have a generalization in the form
\be
^{R}I^{\alpha-2k} f(P) = (-1)^k \, \Delta^k 
\, ^RI^{\alpha} f(P) .
\ee

Note that the value of $m$ can be chosen sufficiently large 
to fulfill all the conditions on the fractional order $\alpha$
in the property (\ref{SemiGr-LP}).


\section{Fractional d'Alembertian for 
pseudo-Euclidean space-time $\mathbb{R}^n_{1,n-1}$}

For the first time  the fractional d'Alembertian 
has been suggested by Riesz in \cite{Riesz} 
(see also \cite{PRTV}).
Note that in the Riesz paper \cite{Riesz}, 
the d'Alembertian $\Box$
is denoted by $\Delta$ and it is called 
for the Laplace operator for $\mathbb{R}^n_{1,n-1}$.
We will use the generally accepted notation.

For the pseudo-Euclidean space-time $\mathbb{R}^n_{1,n-1}$,
we  use
\be
r^2_{QP}=r^2_{PQ} =
(x_1 -y_1)^2 - \sum^n_{j=2} (x_j-y_j)^2 ,
\ee
and the operator
\be
\Box =  \frac{\partial^2}{\partial x^2_1}
- \sum^n_{j=2} \frac{\partial^2}{\partial x^2_j} .
\ee

The Riesz fractional integal is defined as
\be
^RI^{\alpha} f(P) =
\frac{1}{H_n (\alpha)} \int_{\mathbb{R}^n_{1,n-1}}
f(Q) \,r^{\alpha-n}_{PQ} dQ ,
\ee
where $H_n(\alpha)$ is defined 
by the equations (see Eq. 20 and 88 of \cite{Riesz}) 
\be
H_n(\alpha) = 
\pi^{(n-2)/2} 2^{\alpha-1} \, \Gamma (\alpha/2) \,
\Gamma ((\alpha+2-n)/2) = \pi^{(n-1)/2} \, 
\frac{ \Gamma ((\alpha+2-n)/2) \, 
\Gamma (\alpha)}{ \Gamma ((\alpha+1)/2) } .
\ee
This expression of $H_n(\alpha)$
does not coincide with the Euclidean case (\ref{HnE}).
To ensure the convergence of the integral, 
we should assume that the parameter
$\alpha$ satisfies the condition $\alpha-n>-2$,
i.e. $\alpha > n-2$.

There is the following important property
(see Eq. 67 of Chapter 3 in \cite{Riesz})
\be \label{Tru2}
^RI^{\alpha} f(P) = \, ^RI^{\alpha+2m} \, \Box^m f(P) ,
\ee
that allows us to use an analytic continuation of 
$^RI^{\alpha} f(P)$ for negative values of $\alpha$ if
\[ \alpha > n -2 - 2m .\]

Using  the property (\ref{Tru2}), 
we can define the fractional d'Alembertian 
for $n$-dimensional pseudo-Euclidean 
space-time $\mathbb{R}^n_{1,n-1}$ by the equation
\be \label{Box}
\Box^{\alpha/2} f(P) = \,
^RI^{2m-\alpha} \, \Box^m f(P) ,
\ee
where $m > (\alpha+n-2)/2$ and $m \in \mathbb{R}$.

It is important that the semi-group property 
\be \label{SemiGr-DL}
\Box^{\alpha_1/2} \, \Box^{\alpha_2/2} f(P) =
\Box^{(\alpha_1+\alpha_2)/2} f(P) ,
\ee
holds for operators (\ref{Box})
if $0<\alpha_1 <2m+2-n$, $0<\alpha_2 <2m+2-n$, and
$\alpha_1 +\alpha_2 < 2m+2-n$.

The fractional d'Alembertian $\Box^{\alpha/2}$ 
for $n$-dimensional pseudo-Euclidean 
space-time $\mathbb{R}^n_{1,n-1}$
can be considered as an inverse Fourier's integral transform 
${\cal F}^{-1}$ of $|{\bf k}|^{\alpha}$ by 
\be \label{FourBox}
( \Box^{\alpha/2} \varphi)(x)= 
{\cal F}^{-1} \Bigl( |{\bf k}|^{\alpha} ({\cal F} \varphi)
({\bf k}) \Bigr) ,
\ee
where $\alpha > 0$ and $x \in \mathbb{R}^n_{1,n-1}$.


\section{Scalar field in pseudo-Euclidean space-time}


Let us consider  a scalar field $\varphi (x)$
in the $n$-dimensional pseudo-Euclidean space-time 
$\mathbb{R}^n_{1,n-1}$ that is described 
by the field equation
\be \label{31}
\left( \Box + m^2 \right) \varphi (x) = J(x) ,
\ee
where $\Box$ is the d'Alembert operator, 
$\varphi (x)$ is a real field, and $x \in \mathbb{R}^n_{1,n-1}$ 
is the space-time vector with components $x_{\mu}$, 
where $\mu=0,1,2, . . . , n$. 
Suppose that the scalar field $\varphi (x)$ has a source $J(x)$,
and we have put $\hbar=1$.
Field equation (\ref{31}) follows from 
stationary action principle, $\delta S [\varphi]=0$, 
where the action $S [\varphi]$ has the form
\be \label{32}
S [\varphi] = \int_{\mathbb{R}^n_{1,n-1}} 
d^n x \, {\cal L} [\varphi (x)] 
\ee
with the Lagrangian
\be \label{KGE}
{\cal L} [\varphi (x)] =
- \frac{1}{2} \varphi (x) \Bigl( \Box +m^2\Bigr) \varphi (x) .
\ee

The solution of (\ref{KGE}) can be represented in the form
\be
\varphi (x) = - \int \Sigma_F(x-y) \, J(y)\, d^n y ,
\ee
where $\Sigma_F(x-y)$ is the so-called Feynman propagator, obeying 
\be
(\Box +m^2 - i \varepsilon) \, \Sigma_F(x-y) (x) = 
- \delta (x) .
\ee
Here $\delta (x)$ is the Dirac delta-function.
It is easy to see that $\Sigma_F(x-y)$ 
has the Fourier representation 
\be
\Sigma_F(x-y) (x) = 
\int \frac{d^n k}{(2 \pi)^n}  
\frac{e^{-ikx}}{k^2 -m^2 + i\varepsilon} .
\ee



We can consider $\varphi^4$ theory with 
an interaction constant $g$.
It is known that $\Sigma_F (0)$ is a divergent quantity, 
which modifies the free particle propagator and contributes 
to the self-energy. 
In momentum space it corresponds to the loop integral 
\be \label{Ig1}
g \, \Sigma_F (0) =
g \, \int \frac{d^n q}{(2 \pi)^n} \frac{1}{q^2 -m^2} .
\ee
There are $n$th powers of $q$ in the numerator and 
two in the denominator, so 
the integral diverges quadratically at large $q$, i.e. 
it is ultra-violet divergent.
Expression (\ref{Ig1}) corresponds to a one-loop diagram Feynman 
that has the order g. 
Another divergent diagram is the $O(g^2)$ graph. 
The corresponding expression of the
Feynman integral is
\be \label{Ig2}
g^2 \int \frac{d^n q_1}{(2 \pi)^n} \frac{d^n q_2}{(2 \pi)^n}
\frac{\delta(q_1+q_2-p_1-p_2)}{(q^2_1-m^2) (q^2_2 - m^2)} =
g^2 \int \frac{d^n q}{(2 \pi)^n} 
\frac{1}{(q^2-m^2) ((p_1+p_2-q)^2 - m^2)} .
\ee
Here there are $n$th powers of $q$ in numerator and 
the four powers in denominator. 
For $n=4$, we have four powers of $q$ in both numerator and 
denominator, so we get a logarithmic divergence.


\section{Regularization}

Regularization is a method of isolating the divergences. 
There are several techniques of regularization. 
The most intuitive one is to introduce 
a lattice reguariazation \cite{AHEP2014}. 

1. The Pauli-Villars regularization  
modifies the quadratic terms
by subtraction off the same  quadratic terms 
with a much larger mass
\be
\frac{1}{2} \partial_{\mu} \varphi \, 
\, \partial_{\mu} \varphi  - \frac{1}{2} m^2 \ \varphi^2 
- \left( \frac{1}{2} \partial_{\mu} \phi \, 
\, \partial_{\mu} \phi  - \frac{1}{2} M^2 \, \phi^2 \right) .
\ee
The Pauli-Villars regularization is a contribution
of another field (the Pauli-Villars field $\phi$) with the 
same quantum numbers as the original fieeld, 
but having the opposite statistics. 
For example, in the $\varphi^4$-theory,  
the Pauli-Villars field $\phi$ is used, and
the Pauli-Villars regularization changes the propagator as 
\be 
\frac{1}{p^2+m^2}  \ \longrightarrow \
\frac{1}{p^2+m^2} - \frac{1}{p^2+m^2} 
\frac{e^{-p^2/\Lambda^2}}{p^2+M^2} .
\ee

2. The Gaussian cutoff regularization is implemented by 
the modified kinetic term in the form
\be
\frac{1}{2} \partial_{\mu} \varphi \, e^{\Box/\Lambda^2} \,
\partial_{\mu} \varphi  ,
\ee 
which changes the propagator 
\be 
\frac{1}{p^2+m^2}  \ \longrightarrow \
\frac{e^{-p^2/\Lambda^2}}{p^2+m^2} .
\ee

3. The higher derivative regularization 
also modifies the quadratic terms such as
\be
\frac{1}{2} \partial_{\mu} \varphi \, 
\left( 1- \frac{\Box}{\Lambda^2} \right)
\, \partial_{\mu} \varphi  
+ \frac{1}{2} m^2 \
\varphi \, 
\left( 1- \frac{\Box}{\Lambda^2} \right) 
\, \varphi .
\ee 
Then the propagator is modified 
\be
\frac{1}{p^2+m^2}   \ \longrightarrow \
\left(1 + \frac{p^2}{\Lambda^2} \right) 
\frac{1}{p^2+m^2} .
\ee


\section{Regularization by fractional derivatives}

We propose a new regularization, 
where integer-order differential operators are 
replaced by fractional-order operators.
For example, we can replace $\Box$, which is the d'Alembert operator in the field equations, 
by the fractional d'Alembert operator $\Box^{\alpha}$: 
\be
\Box \ \longrightarrow \ \Box^{\alpha/2} .
\ee
Here we use the dimensionless coordinates.

The fractional derivative regularization
modifies the quadratic terms such as
\be
- \frac{1}{2} \varphi (x) \Bigl( \Box +m^2\Bigr) \varphi (x) 
\ \longrightarrow \ 
- \frac{1}{2} \varphi (x) \Bigl( \Box^{\alpha/2} +m^2\Bigr) \varphi (x) .
\ee

Using the fractional d'Alembertian 
for pseudo-Euclidean space-time $\mathbb{R}^n_{1,n-1}$,
we can replace field equation (\ref{31}) by the equation
\be \label{FracEq}
\left( \Box^{\alpha/2} + m^2 \right) \varphi (x) = J(x) ,
\ee
where $\Box^{\alpha}$ is the fractional d'Alembert operator, 
$\varphi (x)$ is a real field, and $x \in \mathbb{R}^n_{1,n-1}$ 
is the space-time vector with components $x_{\mu}$, 
where $\mu=0,1,2, . . . , n$.

The solution of (\ref{FracEq}) is
\be
\varphi (x) = - \int \Sigma_F(x-y) \, J(y)\, d^n y ,
\ee
where $\Sigma_F(x-y)$ is the Feynman propagator, obeying 
\be
(\Box^{\alpha/2} +m^2 - i \varepsilon) \, \Sigma_F(x-y) (x) = 
- \delta (x) .
\ee
Using (\ref{FourBox}), it is easy to see 
that $\Sigma_F(x-y) (x)$  has the Fourier representation 
\be
\Sigma_F (x) = 
\int \frac{d^n k}{(2 \pi)^n}  
\frac{e^{-ikx}}{(k^2)^{\alpha/2}- m^2 + i\varepsilon} .
\ee

In this case, $\Sigma_F (0)$ is a divergent quantity, 
which modifies the free particle propagator and contributes 
to the self-energy. 
In momentum space it corresponds to the loop integral 
\be \label{Ig1b}
g \, \Sigma_F (0) =
g \, \int \frac{d^n q}{(2 \pi)^n} \frac{1}{(q^2)^{\alpha/2} -m^2} .
\ee
There are $n$th powers of $q$ in the numerator and 
$\alpha$ in the denominator.
Another divergent diagram is the $O(g^2)$ graph. 
The corresponding expression of the
Feynman integral is
\[
g^2 \int \frac{d^n q_1}{(2 \pi)^n} \frac{d^n q_2}{(2 \pi)^n}
\frac{\delta(q_1+q_2-p_1-p_2)}{((q^2_1)^{\alpha/2}-m^2) ((q^2_2)^{\alpha/2} - m^2)} =
\]
\be \label{Ig2b0}
= g^2 \int \frac{d^n q}{(2 \pi)^n} 
\frac{1}{(q^2)^{\alpha/2}-m^2}  \cdot
\frac{1}{( (p_1+p_2-q)^2  )^{\alpha/2} - m^2} .
\ee
Here there are $n$th powers of $q$ in numerator and 
the $2\alpha$ powers in denominator.


\section{FD-regularization for massless theory}

For simplification, we will consider a massless 
scalar field theory.
For this case, expression (\ref{Ig2}) with $m=0$ has the form
\be \label{Ig2b}
g^2 \int \frac{d^n q_1}{(2 \pi)^n} \frac{d^n q_2}{(2 \pi)^n}
\frac{\delta(q_1+q_2-p_1-p_2)}{((q^2_1)^{\alpha/2}) ((q^2_2)^{\alpha/2} )} =
g^2 \int \frac{d^n q}{(2 \pi)^n} 
\frac{1}{(q^2)^{\alpha/2}}  \cdot
\frac{1}{( (p_1+p_2-q)^2  )^{\alpha/2}} .
\ee

The denominators in the integrand of (\ref{Ig2})
are combined by using the Feynman's parametric integral formula
\be \label{FF}
\frac{1}{A^a \, B^b} = 
\frac{\Gamma (a+b)}{\Gamma (a) \, \Gamma (b)}
\int^1_0 dz \frac{z^{a-1}(1-z)^{b-1}}{ [Az + B(1-z)]^{a+b}} .
\ee

Changing the variables $q \to p$ and $p_1+p_2 \to q$, 
the integral (\ref{Ig2b})  takes the form
\be \label{Ig2c}
I(q)=g^2 \int \frac{d^n p}{(2 \pi)^n} 
\frac{1}{(p^2)^{\alpha/2}} \cdot \frac{1}{((p-q)^2)^{\alpha/2}} .
\ee
Using formula (\ref{FF}), we get
\be \label{int-z}
\frac{1}{(p^2)^{\alpha/2}} \cdot \frac{1}{((p-q)^2)^{\alpha/2}} 
=
\int^1_0 dz \frac{ z^{\alpha/2-1} \, (1-z)^{\alpha/2-1} }{[p^2-2pq(1-z)+q^2(1-z)]^{\alpha}} .
\ee
As a result, we have
\be \label{Ig2c2}
I(q)=g^2 \int^1_0  z^{\alpha/2-1} \, (1-z)^{\alpha/2-1} \, dz 
\, \int \frac{d^n p}{(2 \pi)^n} 
\frac{1}{[p^2-2pq(1-z)+q^2(1-z)]^{\alpha}} .
\ee

 

Using $p=(p_0,{\bf r})$ and the polar coordinates, 
we get that momentum integral of (\ref{Ig2c2}) is
\[
I[n,\alpha,q] =
\int \frac{d^n p}{(p^2-2pQ+M^2)^{\alpha}} =
\]
\be \label{Ig2d}
= \frac{2 \pi^{(n-1)/2}}{\Gamma((n-1)/2)}
\int^{+\infty}_{-\infty} dp_0 
\int^{\infty}_0 
\frac{r^{n-2} \, dr}{ [p^2_0-r^2-2pQ+M^2]^{\alpha} },
\ee
where
\be
Q = q(1-z) , \quad M^2 = q^2(1-z) .
\ee

This integral is Lorentz invariant, so 
we evaluate it in the frame $q_{\mu}=(\mu,0)$. 
Then $2pQ = 2 \mu (1-z) p_0$. 
Changing variables to $p^{\prime}_{\mu}=p_{\mu}-q_{\mu}(1-z)$, 
which implies that
\be 
p^2_0 - 2 q (1-z) p_0 =
p^2_0 - 2 \mu (1-z) p_0 = (p^{\prime}_0)^2-q^2 (1-z)^2  , 
\ee
we have 
\be \label{Ig2d2}
I[n,\alpha,q] = \frac{2 \pi^{(n-1)/2}}{\Gamma((n-1)/2)}
\int^{+\infty}_{-\infty} dp^{\prime}_0 
\int^{\infty}_0  \frac{r^{n-2} \, dr}{ 
[(p^{\prime}_0)^2-r^2+q^2z(1-z)]^{\alpha} }  ,
\ee
where we use $q^2(1-z)-q^2(1-z)^2=q^2 z(1-z)$.

Using the beta function
\be
2 \int^{\infty}_0 \frac{t^{2x-1}}{(1+t^2)^{x+y}} = B(x,y) =
\frac{\Gamma (x) \, \Gamma (y)}{ \Gamma(x+y)} ,
\ee
we get
\be \label{Int-s1}
\int^{\infty}_0 \frac{s^{\beta} \, ds}{(s^2+M^2)^{\alpha}} =
\frac{\Gamma((1+\beta)/2) \, \Gamma (\alpha -(1+\beta)/2)}{
2(M^2)^{\alpha - (1+\beta)/2} \, \Gamma(\alpha)} ,
\ee
where we can use
\be \label{Mb}
M^2= - (p^{\prime}_0)^2 -q^2 z(1-z) , 
\quad \beta =n-2 .
\ee



Using (\ref{Int-s1}) with (\ref{Mb}), equation (\ref{Ig2d2}) gives
\[
I[n,\alpha,q] = 
\pi^{(n-1)/2} \frac{\Gamma (\alpha-(n-1)/2)}{\Gamma (\alpha)} 
\int^{+\infty}_{-\infty}
\frac{dp^{\prime}_0}{ [ - (p^{\prime}_0)^2 -q^2 z(1-z) ]^{\alpha -(n-1)/2}} =
\]
\[
= (-1)^{ - \alpha +(n-1)/2} \pi^{(n-1)/2} 
\frac{\Gamma (\alpha-(n-1)/2)}{\Gamma (\alpha)} 
\int^{+\infty}_{-\infty}
\frac{dp^{\prime}_0}{ [ (p^{\prime}_0)^2 + q^2 z(1-z) ]^{\alpha -(n-1)/2}} .
\]
Using (\ref{Int-s1}), we get
\be
I[n,\alpha,q] = i \, \pi^{n/2} 
\frac{\Gamma (\alpha-n/2)}{\Gamma (\alpha)} 
\frac{1}{ [ - q^2 z(1-z) ]^{\alpha -n/2}} .
\ee


As a result, we obtain 
\be \label{final-int}
I[n,\alpha,q] =
\int \frac{d^n p}{[p^2-2pq(1-z)+q^2(1-z)]^{\alpha}} =
i \pi^{n/2} \frac{\Gamma (\alpha - n/2)}{\Gamma (\alpha)}
\frac{1}{[- q^2 z (1-z)]^{\alpha-n/2}} .
\ee

Using (\ref{final-int}), expression (\ref{Ig2c2}) has the form
\be \label{Ig2c3}
I(q)= i g^2 
\pi^{n/2} \frac{\Gamma (\alpha - n/2)}{\Gamma (\alpha)}
\int^1_0  
\frac{z^{\alpha/2-1} \, (1-z)^{\alpha/2-1} \, dz }{[- q^2 z (1-z)]^{\alpha-n/2}} .
\ee
Then we have
\be \label{Ig2c22}
I(q)= i g^2 
\pi^{n/2} \frac{\Gamma (\alpha - n/2)}{\Gamma (\alpha)}
\frac{1}{[- q^2 ]^{\alpha-n/2}} 
\int^1_0 z^{(n-\alpha)/2-1} \, (1-z)^{(n-\alpha)/2-1} \, dz .
\ee

For $n=4$ and $\alpha=2+\varepsilon$, we get
\be
\frac{1}{[- q^2 ]^{\alpha-n/2}} =
(-1)^{n/2-\alpha} \left(1 + \varepsilon \, ln (q^2) + 
O(\varepsilon^2) \right) .
\ee
Here we can use
\be
(-1)^{n/2-\alpha} =
(-1)^{- \varepsilon} = e^{i \pi \varepsilon} =
1 + i \pi \varepsilon .
\ee

We can introduce the function
\be \label{123a}
A(n,\alpha) =\int^1_0  
z^{(n-\alpha)/2-1} \, (1-z)^{(n-\alpha)/2-1} \, dz .
\ee
For $n=4$ and $\alpha=2+\varepsilon$, expression (\ref{123a}) takes the form
\be \label{123b}
A(4,2+ \varepsilon) =
\int^1_0 (z(1-z))^{- \varepsilon/2} \, dz 
= 1- \frac{\varepsilon}{2} \, \int^1_0  \ln (z(1-z)) \, dz 
+ O(\varepsilon^2) .
\ee
Using
\be
\int^1_0  \ln (z(1-z)) \, dz = -2 ,
\ee
equation (\ref{123b}) is written as
\be
A(4,2+ \varepsilon) = 1 +\varepsilon + O(\varepsilon^2) .
\ee

For 4-dimensional space time,
\be \label{GG}
\frac{\Gamma (\alpha - n/2)}{\Gamma (\alpha)} =
\frac{1}{(\alpha-1)(\alpha-2)} .
\ee
For $n=4$ and $\alpha = 2+\varepsilon$, equation (\ref{GG}) gives
\be 
\frac{\Gamma (\alpha - n/2)}{\Gamma (\alpha)} =
\frac{1}{(1+\varepsilon) \, \varepsilon} =
\frac{1}{\varepsilon} - 1 .
\ee

Then integral (\ref{Ig2c22}) with
$n=4$ and $\alpha = 2+\varepsilon$, has the form
\be \label{Ig2c32}
I(q)= i g^2 \pi^{2} \left( \frac{1}{\varepsilon} - 1 \right) \,
(1 +\varepsilon ) \,
(1 + i \pi \varepsilon) \, \left(1 + \varepsilon \, ln (q^2) \right).
\ee

As a result, we get
\be \label{Ig2c33}
I(q)= \frac{ i g^2 \pi^{2}}{\varepsilon} +
i g^2 \pi^{2} \ln(q^2) - \pi^3 g^2 + O(\varepsilon) .
\ee

\newpage

{\bf Remarks}

For the special cases, we have
\be
A(2,1)= \pi , \quad A(4,1) = \frac{\pi}{8} , \quad
A(4,2)= 1 , \quad A(3,1)=1 .
\ee

The well-known asymptotic expression 
for the gamma-function is
\be
\Gamma (\varepsilon) = \frac{1}{\varepsilon}
-\gamma+ \frac{1}{2} \left( \gamma^2 +\frac{\pi^2}{6} \right) \, \varepsilon + O(\varepsilon^2)  \quad (\varepsilon \to 0).
\ee

For 2-dimensional space-time,
\be
\frac{\Gamma (\alpha - n/2)}{\Gamma (\alpha)} =
\frac{1}{\alpha-1} .
\ee
If $\alpha = 1+\varepsilon$, then
\be
\frac{\Gamma (\alpha - n/2)}{\Gamma (\alpha)} =  
\frac{1}{\varepsilon} .
\ee

For 4-dimensional space time (n=4),
\be
\frac{\Gamma (\alpha - n/2)}{\Gamma (\alpha)} =
\frac{1}{(\alpha-1)(\alpha-2)} .
\ee
If $\alpha = 1+\varepsilon$, then
\be
\frac{\Gamma (\alpha - n/2)}{\Gamma (\alpha)} =
\frac{1}{ \varepsilon \, (\varepsilon-1) } =
-\frac{1}{\varepsilon} - 1 .
\ee
If $\alpha = 2+\varepsilon$, then
\be
\frac{\Gamma (\alpha - n/2)}{\Gamma (\alpha)} =
\frac{1}{(1+\varepsilon) \, \varepsilon} =
\frac{1}{\varepsilon} - 1 .
\ee


{\bf Remark}. By changing variables to 
$p^{\prime} = p - q(1- z)$, 
we see that the denominator in the integrand 
of equation (\ref{int-z}) is the square of 
$(p^{\prime})^2 - m^2 + q^2 z(1 - z)$. 
But $d^n p^{\prime} = d^n p$, so relabelling $p^{\prime} \to p$, 
expression (\ref{Ig2b}) becomes 
\be \label{Ig2c4}
I(q)=g^2 \int^1_0  z^{\alpha/2-1} \, (1-z)^{\alpha/2-1} \, dz 
\, \int \frac{d^n p}{(2 \pi)^n} 
\frac{1}{[p^2+q^2 z (1-z)]^{\alpha}} . 
\ee


 
\section*{Conclusion}

In this paper, we proposed a new regularization
of quantum field theories, which is based on
the Riesz fractional derivatives of 
non-integer order $\alpha>0$.
The fractional derivatives of non-integer order 
are characterized by nonlocality, 
the proposed FD regularization is actually 
a regularization by space-time nonlocality.

Note that the Riesz fractional derivatives of 
non-integer order \cite{Riesz,PRTV} can be used not only for 
FD regularization.
An application of fractional calculus in quantum field theory
allows us to take into account space and time nonlocality 
\cite{Efimov1967,Efimov1985,Lim-1,Lim-3,AHEP2014}.
It should be noted that we can use the Riesz fractional derivatives \cite{Riesz,PRTV},
the fractional Laplacian and d'Alembertian in the lattice 
quantum field theories \cite{AHEP2014}.
Using the exact discretization of the fractional Laplacian d'Alembertian \cite{CMA2017},
we can consider exact discrete analogs \cite{CNSNS2016-2} of 
the quantum field theories.
The fractional derivatives of the Riesz type \cite{Riesz,PRTV} 
are directly connected to 
the long-range properties and nonlocal interactions
\cite{JPA2014,AMC2015,CNSNS2016-2}. 
Because of this, this tool can be used not only for regularization, 
but also for constructing new quantum theories in 
high-energy physics and condensed matter physics.

 
\section*{Appendix}

For the fractional case, we should consider the integrals
\be \label{Int-s2} 
K[\alpha , \beta, \gamma, M^2] =
\int^{\infty}_0 \frac{s^{\gamma} \, ds}{(s^{\alpha}+M^2)^{\beta}} 
\ee
instead of the integrals of equation (\ref{Int-s1}).
Using new variable $z$ such that $z^2=s^{\alpha}$, and
\be
z =s^{\alpha/2} , \quad s=z^{2/\alpha} ,
\ee
\be
ds = (2/\alpha) \, z^{2/\alpha-1} \, dz , 
\quad
s^{\gamma} = z^{2 \gamma /\alpha} ,
\ee
equation (\ref{Int-s2}) gives 
\be
\frac{2}{\alpha}
K[\alpha , \beta, \gamma, M^2] =
\int^{\infty}_0 dz \frac{z^{ 2(\gamma+1)/ \alpha-1 } }{(z^2+M^2)^{\beta}} .
\ee
Then using equation (\ref{Int-s1}), we obtain
\be
K[\alpha , \beta, \gamma, M^2] =
\frac{ 2\, \Gamma((\gamma+1)/ \alpha) \, 
\Gamma (\beta -(\gamma+1)/ \alpha)}{ 2 \, \alpha \,
(M^2)^{\beta - (\gamma+1)/ \alpha } \, \Gamma(\beta)} .
\ee


\section*{Conflict of Interests}

The author declares that there is no conflict of interests regarding the publication of this paper.


\section*{Acknowledgments}

This work was partially supported by grant NSh-7989.2016.2



\end{document}